\pdfoutput=1
\documentclass{article}
\usepackage{spconf,amsmath,graphicx,amsfonts}
\usepackage{booktabs}
\usepackage{diagbox} 
\usepackage{multirow} 
\usepackage{makecell} 
\usepackage{longtable} 

\usepackage[colorlinks,linkcolor=black,urlcolor=blue,citecolor=black]{hyperref}

\title{Streaming Voice Conversion Via Intermediate Bottleneck Features And Non-streaming Teacher Guidance}
\name{\makecell{Yuanzhe Chen$^*$\thanks{* chenyuanzhe@bytedance.com }, Ming Tu, Tang Li, Xin Li, Qiuqiang Kong, Jiaxin Li, Zhichao Wang\\  Qiao Tian, Yuping Wang, Yuxuan Wang}}  
\address{Speech, Audio \& Music Intelligence (SAMI), ByteDance}
\begin{document}
%
\maketitle
\begin{abstract}
Streaming voice conversion (VC) is the task of converting the voice of one person to another in real-time. Previous streaming VC methods use phonetic posteriorgrams (PPGs) extracted from automatic speech recognition (ASR) systems to represent speaker-independent information.
However, PPGs lack the prosody and vocalization information of the source speaker, and streaming PPGs contain undesired leaked timbre of the source speaker. In this paper, we propose to use intermediate bottleneck features (IBFs) to replace PPGs. VC systems trained with IBFs retain more prosody and vocalization information of the source speaker. Furthermore, we propose a non-streaming teacher guidance (TG) framework that addresses the timbre leakage problem. Experiments show that our proposed IBFs and the TG framework achieve a state-of-the-art streaming VC naturalness of 3.85, a content consistency of 3.77, and a timbre similarity of 3.77 under a future receptive field of 160 ms which significantly outperform previous streaming VC systems.

\end{abstract}
\section{Introduction}
\label{sec:intro}


Voice conversion (VC) is the task of converting the voice of one person to another while keeping the content consistent. VC has a strong demand for audio editing and voice interaction scenarios. In the past, many researches focused on non-streaming VC, such as AutoVC \cite{qian2019autovc}. Non-streaming VC methods require users to enter a complete voice sentence before returning the result. Different from non-streaming VC, streaming VC is a task to continuously convert the voice of one person to another in real time. In recent years, streaming VC has many applications such as livestreaming, avatar, and real-time communication (RTC). 


Due to the difficulty of acquiring parallel corpus containing paired utterances between different speakers, current VC methods are mostly based on non-parallel corpus such as auto-encoder based VC \cite{qian2019autovc, nguyen2022nvc}, phonetic posteriorgrams (PPGs)  based VC \cite{2016Phonetic, zhao2022disentangling}, and self-supervised representation based VC \cite{choi2021neural}. Those methods use an encoder to extract speaker-independent representations, such as the content of the input speech and use a decoder along with the input speaker ID as a condition to reconstruct the input audio. 
Recent streaming VC methods include using streaming automatic speech recognition (ASR) encoder to extract PPGs \cite{ronssin2021ac, chen2022streaming} and designing causal model structures for VC 
\cite{kameoka2021fasts2s, tobing2021low, hayashi2022investigation}. 

A perfect representation extracted from the encoder should contain the complete content and prosody but not the timbre of the source speaker. However, the receptive fields of streaming ASR systems are limited to the past, so the recognition accuracy of streaming ASR systems are worse than non-streaming ASR systems \cite{zhang2020unified}. Second, PPGs only represent finite phoneme classes and lack paralinguistic information such as prosody and vocalization \cite{web}. Third, due to the streaming ASR encoder having limited a receptive field, the streaming ASR encoder contains the undesired timbre of the source speaker, which is referred to as timbre leakage. In inference, timbre leakage will result in the VC output containing timbres of both the source and target speakers.

To address those problems, we first propose to use intermediate bottleneck features (IBFs) to replace PPGs as source speaker representations to improve the content robustness of the VC system. IBFs contain more fine-grained pronunciation and paralinguistic information than PPGs. IBFs can significantly improve the pronunciation accuracy, vocalization, and prosody retention in both non-streaming VC and streaming VC systems. Second, we propose a non-streaming teacher guidance (TG) framework which uses a pre-trained non-streaming VC model as a teacher to generate parallel training data of different source speakers and target speakers. Then, we propose a student model to learn from those parallel data instead of using self-reconstruction loss. We analyze that the TG framework will force the streaming VC system to \textit{ignore} the timbre of the source speaker that significantly mitigate the timbre leakage problem.


The rest of this paper is organized as follows. Section~\ref{sec:ibf} introduces our proposed IBFs based VC system. Section~\ref{sec:tg} introduces our proposed non-streaming TG training strategy. Section~\ref{sec:exp} shows experiments. Section~\ref{sec:conclusion} concludes this paper.


\section{Voice Conversion with Intermediate Bottleneck Features}
\label{sec:ibf}

Both mainstream non-streaming and streaming VC systems \cite{2016Phonetic, chen2022streaming} consist of three parts. The first part is a pre-trained ASR Encoder to extract PPGs. The second part is an acoustic model (AM) taking PPGs and the target speaker ID as input to predict the mel spectrum of the target speaker. The third part is a vocoder for synthesizing the waveform from the mel spectrum. 



\subsection{Intermediate Bottleneck Features}
\label{ssec:asr}

PPGs have the disadvantage of only containing the content but not the paralinguistic information of input speech, such as prosody and vocalization. To address this problem, we propose to use IBFs extracted from an ASR sub-encoder instead of PPGs as the content representation.
Different layers in an ASR encoder contain different level of information. Shallower layers contain more low-level timbre information, while deeper layers contain more high-level content information \cite{li2020does}. 
Previous PPGs-based VC methods only use the last layer output of the ASR encoder. Therefore, PPGs lack low-level details, such as prosody, breathing, and whispers. 
For an ASR encoder with $K$ layers, we denote the ASR sub-encoder as the first $k$ layers of the ASR encoder where $ k < K $. We denote the output of the ASR sub-encoder as IBFs. The PPGs are equivalent to the IBFs when $k = K$. 


\subsection{Acoustic Model}
\label{ssec:am}
\begin{figure}[t]
\begin{minipage}[b]{1.0\linewidth}
  \centering
  \centerline{\includegraphics[width=8.5cm]{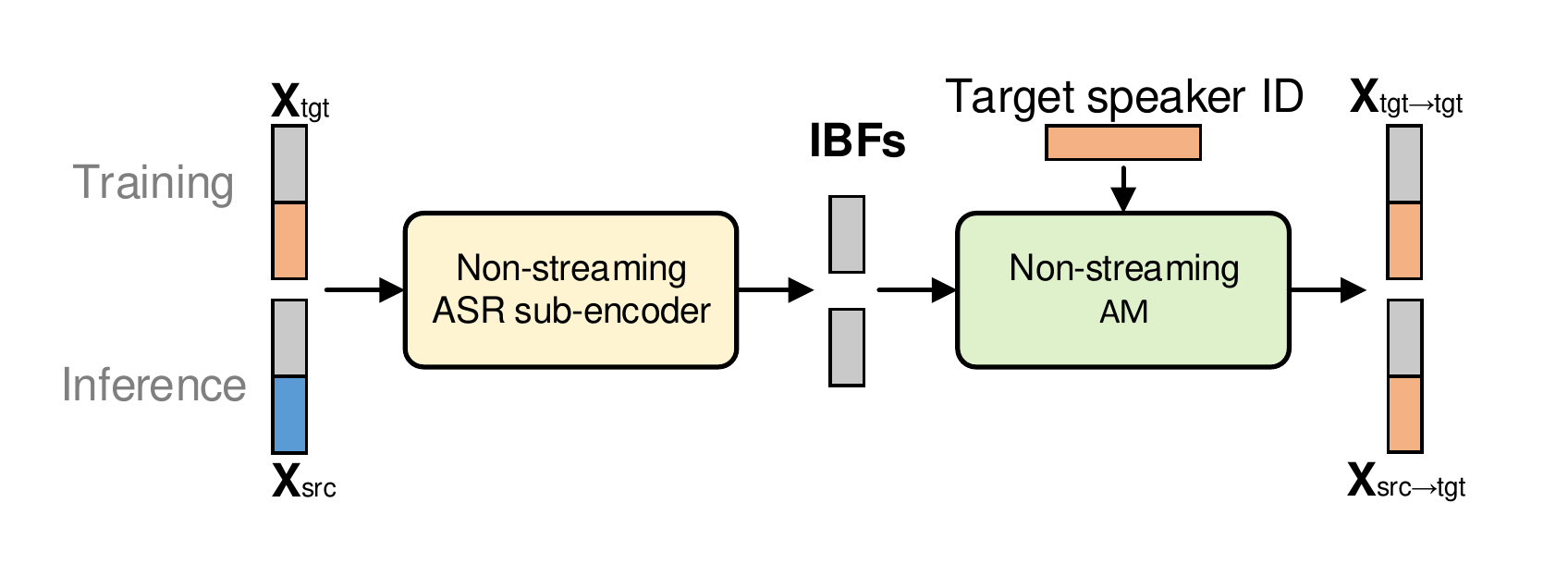}}
 \vspace{-10pt}
  \centerline{(a) Non-streaming + IBFs + reconstruction (one-to-one)}\medskip
  \vspace{-5pt}
\end{minipage}

\begin{minipage}[b]{1.0\linewidth}
  \centering
  \centerline{\includegraphics[width=8.5cm]{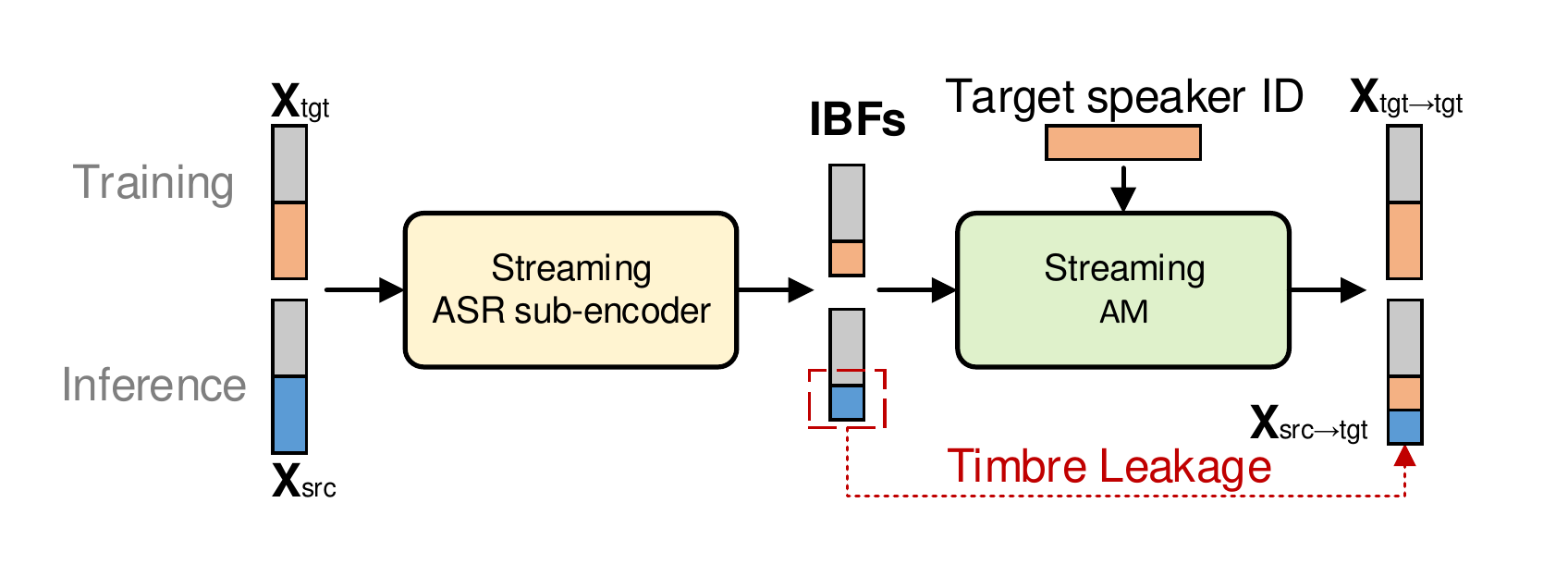}}
  \vspace{-10pt}
  \centerline{(b) Streaming + IBFs + reconstruction (one-to-one)}\medskip
  \vspace{-5pt}
\end{minipage}

\begin{minipage}[b]{1.0\linewidth}
  \centering
  \centerline{\includegraphics[width=8.5cm]{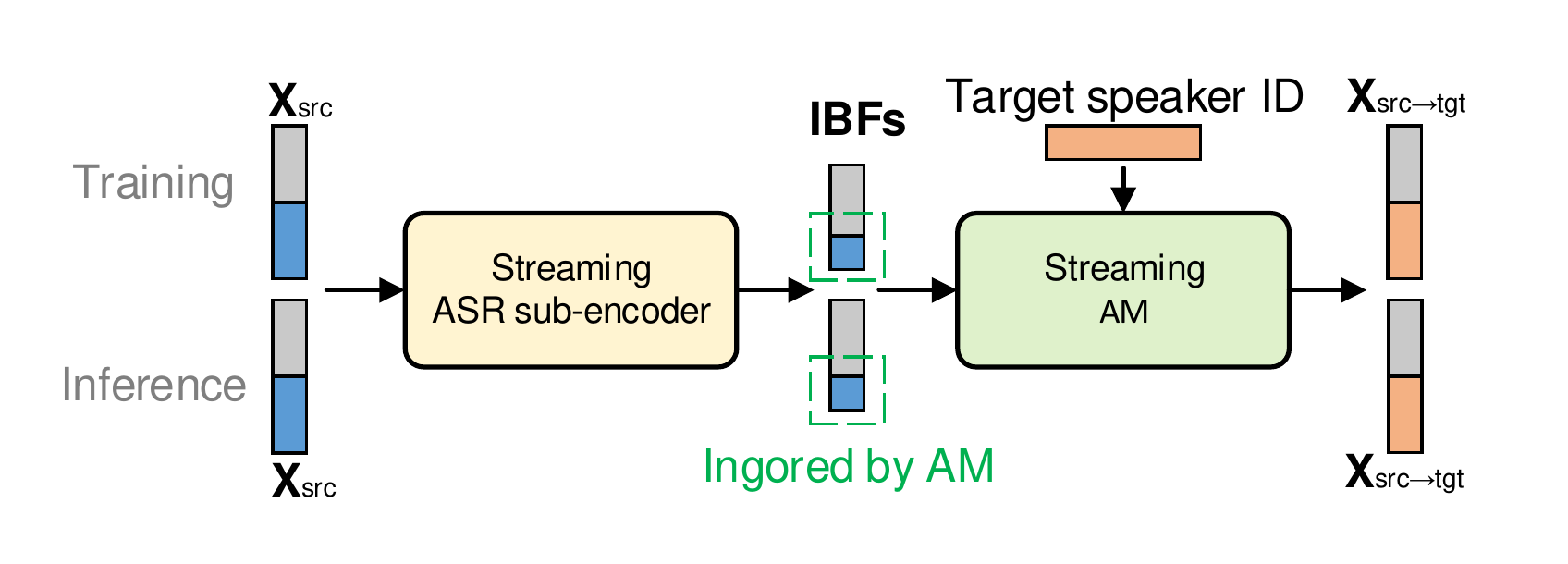}}
  \vspace{-5pt}
  \centerline{(c) Streaming + IBFs + TG (many-to-one) }\medskip
  \vspace{-5pt}
\end{minipage}

\vspace{-5pt}
\caption{
Training and inference of non-streaming and streaming VC systems.
Gray bars represent content information. Blue and orange bars represent the source speaker and target speaker timbre information, respectively}
\label{fig:why}
\end{figure}

We use an AM to predict the target speaker mel spectrum from the source speaker IBFs and the target speaker ID. For non-streaming VC, the AM is a 6-layer conformer encoder-decoder structure similar to DelightfulTTS \cite{liu2021delightfultts}. For streaming VC, the AM consists of a chunk-level encoder and an autoregressive LSTM frame-level decoder 
for streaming inference.
The training and inference process of IBFs-based VC can be formalized by:
\begin{equation}
X_{\text{src} \rightarrow \text{tgt}} =M(E(X_{\text{src}}), s_{\text{tgt}}),
\label{eq:acoustic_model}
\end{equation}
where $X_{\text{src}}$ is log mel spectrum of the input speech and $ E(\cdot) $ is the sub ASR encoder to extract IBFs. The vector $ s_{\text{tgt}} $ represents the target speaker ID. The acoustic model $M(\cdot,\cdot)$ takes the source speaker IBFs and the target speaker ID to predict the target speaker mel spectrum $X_{\text{src} \rightarrow \text{tgt}}$. 
In training, we use $ X_{\text{src}} = X_{\text{tgt}} $ and a self-reconstruction loss to train the VC system:
\begin{equation}
L_{\text{recon}}=\left \| X_{\text{tgt} \rightarrow \text{tgt}} - X_{\text{tgt}} \right \|_1.
\label{eq:reconstruct}
\end{equation}
\noindent In addition, we add a LSGAN \cite{mao2017least} regularization term to (\ref{eq:reconstruct}) to avoid the over-smoothing problem. In inference, we input the source speaker mel spectrum $ X_{\text{src}} $ and the target speaker ID $ s_{\text{tgt}} $ to (\ref{eq:acoustic_model}) to predict the converted voice $ X_{\text{src} \rightarrow \text{tgt}} $. Fig.~\ref{fig:why}(a) shows the training and inference of the non-streaming VC system.



\subsection{Vocoder}
\label{ssec:vocoder}
We apply a TFGAN \cite{tian2020tfgan} vocoder to synthesis mel spectrums into waveforms. The TFGAN vocoder consists of upsampling layer and residual one-dimensional convolutional layers. 
For streaming inference, we limit the kernel size of each convolutional layer in the vocoder such that the prediction for each frame depends only on short-term future inputs.

\begin{figure*}[h]
    \vspace{-10pt}
  \centering
  \includegraphics[width=1\linewidth]{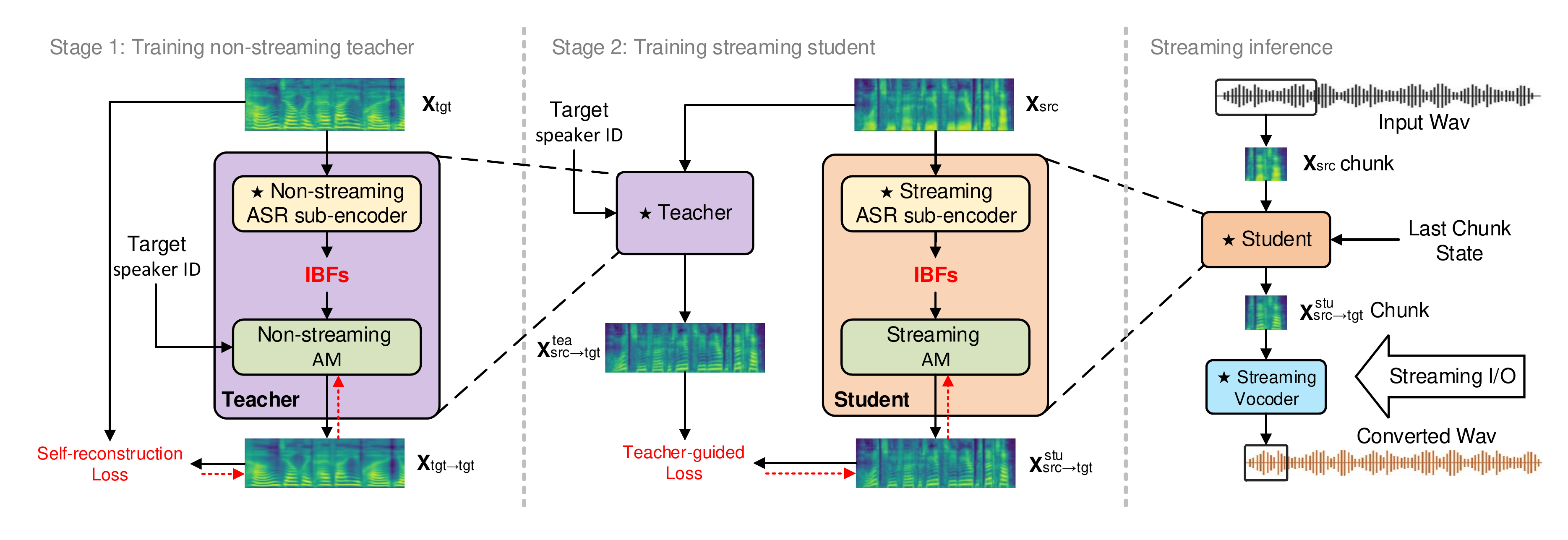}\vspace{-15pt}
  \caption{The overall TG framework. The stars indicate pre-trained and frozen modules. Red arrows show gradient flows. 
  }\label{framework}\vspace{-5pt}
\end{figure*}

\section{Non-streaming Teacher Guidance}
\label{sec:tg}
Previous non-streaming VC systems trained by the self-reconstruction loss (\ref{eq:reconstruct}) have the timbre leakage problem. To explain, streaming ASR systems have limited receptive fields which only use history information. Therefore, streaming ASR systems have reduced performance compared to non-streaming ASR systems which will lead to the PPGs and IBFs extracted from the streaming ASR systems containing undesired timbre of the source speaker which is referred to as the timbre leakage problem, as shown in Fig.~\ref{fig:why}(b). 

To address the timbre leakage problem, we propose a teacher guidance training framework to train the streaming VC system. The core idea is to train a teacher non-streaming VC system to generate parallel data to train the student streaming VC system. Fig.~\ref{fig:why}(c) shows our proposed TG training. The streaming VC learns to \textit{ignore} the leaked timbre of the source speaker in the PPGs or IBFs to achieve a perfect conversion of the target speaker. Fig.~\ref{framework} shows our proposed TG framework, which consists of a non-streaming VC teacher training stage and a streaming VC student training stage.


\subsection{Stage 1: Training Non-streaming Teacher}
\label{subsec:stage1}

Since the non-streaming ASR sub-encoder can extract the content and remove the timbre information of the source speaker well as shown in Fig.~\ref{fig:why}, we train a non-streaming VC model as a teacher to generate high-quality target speaker mel spectrums (\ref{eq:acoustic_model}):
\begin{equation}
X_{\text{src} \rightarrow \text{tgt}}^{\text{tea}} =M^{\text{tea}}(E^{\text{nonstream}}(X_{\text{src}}), s_{\text{tgt}}),
\label{eq:teacher_AM}
\end{equation}
\noindent The pre-trained non-streaming ASR sub-encoder is denoted as $ E^{\text{nonstream}} $. The teacher acoustic model is denoted as $ M^{\text{tea}} $ and $ X_{\text{src} \rightarrow \text{tgt}}^{\text{tea}} $ is the converted mel spectrum. The teacher module is trained by the self-reconstruction loss (\ref{eq:reconstruct}). The left column of Fig.~\ref{framework} shows the training of the teacher module.

\subsection{Stage 2: Training Streaming Student }
Similar to the teacher non-streaming VC module, we denote the student streaming VC model as:
\begin{equation}
X_{\text{src} \rightarrow \text{tgt}}^{\text{stu}} =M^{\text{stu}}(E^{\text{stream}}(X_{\text{src}}), s_{\text{tgt}}),
\label{eq:student_AM}
\end{equation}
\noindent where $ E^{\text{stream}} $ is a pre-trained streaming ASR sub-encoder. The student acoustic model is denoted as $ M^{\text{stu}} $ and $ X_{\text{src} \rightarrow \text{tgt}}^{\text{stu}} $ is the converted mel spectrum. Different from the teacher module, we propose to use generated parallel data and a TG loss instead of the self-reconstruction loss (\ref{eq:reconstruct}) to train the student module:



\begin{equation}
L_{TG}=\| X_{\text{src} \rightarrow \text{tgt}}^{\text{stu}}- X_{\text{src} \rightarrow \text{tgt}}^{\text{tea}} \|_1.
\label{con:t6}
\end{equation}

\noindent Fig.~\ref{fig:why}(c) shows that the source and target speakers are different in both training and inference. The streaming VC system trained with the TG loss learns to ignore the source speaker timbre. The middle column of Fig.~\ref{framework} shows the training of the student module. 


\subsection{Streaming Inference}
The streaming inference stage is shown in the right column of Fig.~\ref{framework}. An audio chunk is sent to the streaming VC system. The first forward propagation starts when the input speech duration is buffered to the chunk size of $E^{\text{stream}}$. 
Then, a streaming vocoder is applied on the predicted mel spectrum to synthesize the waveform of the target speaker.
\vspace{-10pt}





\section{Experiements}
\label{sec:exp}

\subsection{Dataset and Experimental Setup}
\label{ssec:subhead}

 Both non-streaming and streaming ASR systems are pre-trained on 10,000 hours of in-house Mandarin corpus. We use the Aishell-3 \cite{shi2020aishell} corpus containing 218 speakers to train the AM in Section \ref{ssec:am} and the vocoder in Section \ref{ssec:vocoder}. 
 We use a male and a female as target speakers from an in-house dataset, each with 3 hours of audio to finetune the self-reconstruction based AM and the vocoder. All training audio recordings were downsampled to 24 kHz. 
We train a 18-layer conformer-based \cite{gulati2020conformer} and a 18-layer streaming conformer-based \cite{zhang2020unified} ASR encoder  for non-streaming and streaming ASR systems, respectively. 
The chunk size of the streaming ASR encoder is 160 ms. The streaming AM takes IBFs of a chunk as input and outputs the mel spectrum of a chunk. For fair evaluation, both non-streaming and streaming AMs use streaming vocoders to restore the waveform. 
The latency of the entire streaming VC pipeline is about 270ms on a single Intel i5-6267U CPU core. The demos of the VC results of our system can be listened at \href{https://qq547276542.github.io/StreamingVC.github.io/}{https://qq547276542.github.io/StreamingVC.github.io/}.



\begin{figure}
    \vspace{-10pt}
  \centering
  \centerline{\includegraphics[width=10cm]{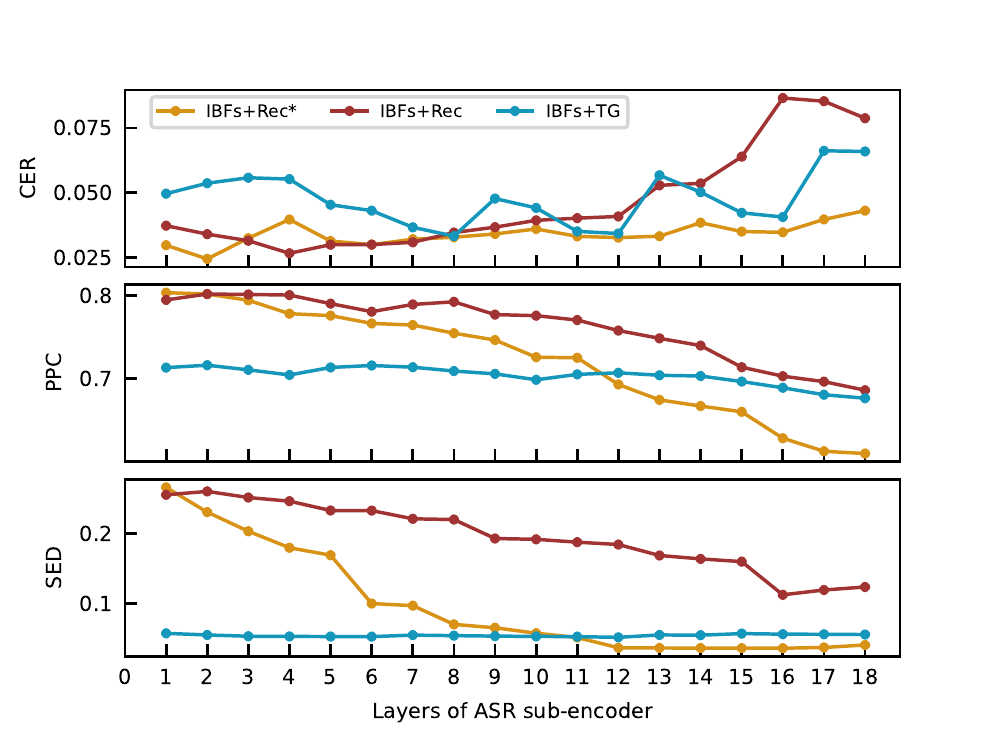}}\vspace{-10pt}
  \caption{CER, PPC, and SED metrics of different VC systems trained with different IBF layers. 
  }\label{layer_ana}\vspace{-5pt}
\end{figure}

\subsection{Objective Evaluation}
\label{ssec:objective}

We use the character error rate (CER) of the predicted waveform and the pitch Pearson correlation (PPC) between the predicted waveform and the corresponding input speech \cite{wang2021enriching} to measure the \textit{content consistency} between the VC output and the input speech. Smaller CER indicates better VC pronunciation result. Larger PPC indicates more prosody information of the input waveform are retained. We use speaker embedding distance (SED) extracted by a pre-trained ECAPA-TDNN \cite{desplanques2020ecapa} to evaluate the timbre similarity between the VC result and the target speech.

Fig.~\ref{layer_ana} shows the the VC results trained with different IBFs layers to explore the optimal layer selection. \textit{IBFs+Rec*} indicates the IBFs-based non-streaming VC model trained with the self-reconstruction loss. The asteroid symbol represents non-streaming systems. \textit{IBFs+Rec} and \textit{IBFs+TG} are streaming VC systems trained with self-reconstruction and TG losses, respectively. For all experiments, The PPG features are equivalent to the 18-layer IBFs. Fig~\ref{layer_ana} shows that lower IBFs results in higher pronunciation accuracy and more prosody information retention than the PPG feature. However, the high SED in low level IBFs indicates that input speech is not converted to the target speaker. This is because shallow IBFs are close to the original input mel spectrum and the self-reconstruction training (\ref{eq:reconstruct}) will cause AM degenerates into identity mapping from the input mel spectrum to the output mel spectrum. 
Fig~\ref{layer_ana} shows that the SED of the IBFs+Rec* system reaches the lowest from the 12th layer, indicating that the 12-layer IBFs are the optimal representations for non-streaming VC.
Fig~\ref{layer_ana} shows that the IBFs+TG system (\ref{eq:teacher_AM}) significantly outperform the IBFs+Rec system (\ref{eq:reconstruct}) in SED, indicating that the TG training overcomes the timbre leakage problem.

\begin{table}[t]
  \begin{center}

\caption{MOS results with 95\% confidence interval.
}   \label{tab:mos}
\resizebox{\columnwidth}{!}{%
\begin{tabular}{lccc}
    \toprule
\textbf{Methods} & \textbf{Naturalness} & \textbf{\makecell{Content \\ Consistency}}   & \textbf{\makecell{Timbre \\ Similarity}} \\  \midrule
Copysyn  & 4.10$\pm$0.03  & -  & 4.07$\pm$0.04 \\
\hline
PPGs+Rec* \cite{2016Phonetic}&   3.91$\pm$0.03 &   3.57$\pm$0.04 & \textbf{3.94$\pm$0.04} \\
IBFs+Rec* &  \textbf{3.93$\pm$0.03}&   \textbf{3.77}$\pm$0.03 & 3.80$\pm$0.04 \\
\hline
PPGs+Rec \cite{ronssin2021ac} &   3.71$\pm$0.04 &   3.69$\pm$0.03 & 3.64$\pm$0.04 \\
PPGs+Rec+Adv \cite{chen2022streaming}&   3.71$\pm$0.05 &   3.67$\pm$0.04 & 3.65$\pm$0.04 \\
IBFs+Rec  &   2.90$\pm$0.06 &   3.72$\pm$0.05 & 2.18$\pm$0.05 \\
PPGs+TG  &   3.84$\pm$0.03 &   3.69$\pm$0.03 & 3.74$\pm$0.04 \\
IBFs+TG &   \textbf{3.85}$\pm$0.03 &   \textbf{3.77$\pm$0.04} & \textbf{3.77}$\pm$0.04 \\
    \bottomrule
\end{tabular}}
\end{center}
\vspace{-0pt}
\end{table}

\subsection{Subjective Evaluation}

\label{ssec:subjective}
We evaluate the mean opinion scores (MOS) of the VC results from the naturalness, content consistency, and timbre dimensions. The test set consists of 100 sentences pronounced by different speakers. We assign 20 reviewers to score the VC results on a scale of 1-5.


Table~\ref{tab:mos} shows the MOS scores of our proposed systems compared with previous systems. The \textit{Copysyn} is referred to as a perfect system that applies a vocoder on the ground truths mel spectrum. The asteriod symbol indicates non-streaming VC systems. The third row shows that the \textit{IBFs-Rec*} non-streaming VC system achieves a naturalness of 3.93 and a content consistency of 3.77, higher than the \textit{PPGs-Rec*} system of 3.91 and 3.57, respectively. The fourth to the eigth rows show the streaming VC results. Previous \textit{PPGs+Rec} only uses the self-reconstruction loss. \textit{PPGs+Rec+Adv} \cite{chen2022streaming} proposed to add an speaker adversarial loss to remove residual timbre of the source speaker. Without TG training, the \textit{IBFs+Rec} system only achieves a naturalness of 2.90, a content consistency of 3.72, and a timbre similarity of 2.18. Our proposed \textit{PPGs+TG} and \textit{IBFs+TG} systems significantly outperform previous systems in all naturalness, content consistency, and timbre similarity. The \textit{IBFs+TG} system achieves a state-of-the-art naturalness, content consistency, timbre similarity of 3.85, 3.77, and 3.77, respectively. The MOS scores of our proposed streaming VC systems are comparble to the best non-streaming \textit{IBFs+Rec*} VC system.



\section{Conclusions}
\label{sec:conclusion}
In this paper, we proposed an robust streaming VC framework. We first propose to replace PPGs with IBFs, thereby significantly improving the pronunciation accuracy, vocalization retention and prosody retention of the VC model. Then we propose a TG training framework instead of reconstructed training to ensure the timbre stability of streaming VC. In the future, we will further explore how to better utilize non-streaming models to guide streaming models in VC tasks.





\bibliographystyle{IEEEbib}
\bibliography{reference}

\end{document}